# SuperNANO: Enabling Nano-Scale Laser anti-counterfeiting Marking and Precision Cutting with Super-Resolution Imaging

Yiduo Chen [1], Bing Yan [1,2], Liyang Yue [1], Charlotte L Jones [1,*] and Zengbo Wang [1,*]

[1] School of Computer Science and Engineering, Bangor University, Dean Street, Bangor, Gwynedd LL57 1UT, UK.
[2] School of Natural and Computing Sciences, Aberdeen University, Aberdeen, AB24 3FX, UK.
* Correspondence: z.wang@bangor.ac.uk.

**Abstract:** In this paper, we present a unique multi-functional super-resolution instrument, the SuperNANO system, which integrates real-time super-resolution imaging with direct laser nanofabrication capabilities. Central to the functionality of the SuperNANO system is its capacity for simultaneous nanoimaging and nanopatterning, enabling the creation of anti-counterfeiting markings and precision cutting with exceptional accuracy. The SuperNANO system, featuring a unibody superlens objective, achieves a resolution ranging from 50 to 320 nm. We showcase the instrument's versatility through its application in generating high-security anti-counterfeiting features on an aluminum film. These 'invisible' security features, which are nanoscale in dimension, can be crafted with arbitrary shapes at designated locations. Moreover, the system's precision is further evidenced by its ability to cut silver nanowires to a minimum width of 50 nm. The integrated imaging and fabricating functions of the SuperNANO make it a pivotal tool for a variety of applications, including nano trapping, sensing, cutting, welding, drilling, signal enhancement, detection, and nano laser treatment.

## 1. Introduction

Laser technology has become indispensable in micro/nano-patterning due to its widespread usage. This versatile tool facilitates the precise creation of intricate structures through non-contact and maskless laser direct-writing techniques. However, a significant challenge in laser processing persists because of optical diffraction limit, particularly when aiming the production of extremely small features [1]. The minimum size of features achievable with surface patterning technologies, such as photolithography and direct laser writing, as well as the resolution of optical imaging systems, is inherently constrained by this limit. Recently, there has been a notable surge in interest in using dielectric microspheres as near-field lenses for super-resolution nano imaging and fabrication. This interest stems from the optical phenomenon known as the photonic nanojet, which aids in focusing laser beams to surpass the diffraction limit, thereby enabling finer resolution in imaging and fabrication processes [2,3]. Over the past few years, superlenses made from dielectric materials, especially in spherical shapes, have emerged as a powerful platform for challenging the diffraction limit [4–6]. These advancements mainly pursue two research directions: firstly, focusing laser beams onto the sample surface to achieve nanopatterning with sub-100 nm feature sizes, and secondly, manipulating light reflected from the sample surface, forming the foundation of super-resolution optical imaging to observe subwavelength objects and structures with resolution as fine as 45 nm [7,8]. Previous publications have demonstrated 80-nm-resolution in laser nanopatterning [9] and 45-50 nm resolution in nanoimaging [7,10], alongside other notable developments [11].

The dual functional nature of the proposed system roots from the distinct particle-lens setup and super-resolution mechanism it employs. In nanopatterning experiments, it is well established that the patterning resolution is proportional to the particle size. Previous researchers have demonstrated the realization of patterning resolutions ranging from 1 μm to sub-micron levels, including sub-100 nm, using relatively small SiO2 particle lenses [12,13]. This mechanism differs from that employed in nano-imaging. In super-resolution imaging, the microsphere superlens operates in a virtual imaging mode. It collects and transforms near-field evanescent waves, which carry high-spatial-frequency information of the object, into propagating waves that reach the far-field by forming a magnified virtual image. Here, the efficiency of evanescent-to-propagating-conversion (ETPC) determines the final imaging resolution, rather than the spot size on the sample surface as in nanopatterning [14,15]. This mechanism enables the use of larger microspheres as superlenses for super-resolution imaging. At present, BaTiO3 (BTG) microspheres with sizes ranging between 20 and 60 μm are widely used as dielectric superlens [16].

Integrating both nanopatterning and nano-imaging capabilities within a single system requires design of the SuperNano objective to accommodate both working modes: focusing mode and virtual imaging mode, normally based on microsphere material, size, refractive index contrast, integration with external components and optimization [17–20]. Additionally, the system possesses the capability to position the microsphere superlens precisely at desired locations and scan over an area, which is crucial for practical applications. Since a single microsphere has a limited field of view (FOV). Consequently, scanning is utilized to expand the FOV of the system for a dynamic imaging [21,22]. Various scanning schemes were used in this field, including integration with an atomic force microscope (AFM) system, encapsulation of the microsphere in a solid film, optical trapping, etc. The most recent scanning design involves bonding the microsphere superlens directly with an objective lens to create a unibody design [23,24]. In 2020, two approaches to unibody design and their applications in nanopatterning [25] and super-resolution imaging [26] were respectively published. The Plano-Convex-Microsphere (PCM) design stands out as a key innovation. It refers a compound-lens design realized by positioning a high-index microsphere onto a Plano-Convex lens and then integrating them into a conventional objective lens. However, these microsphere-based imaging and patterning systems were not integrated with each other, making real-time simultaneous nano-imaging and nano-patterning impossible. This significantly limits its applications in advanced anti-counterfeiting marking and other potential uses, such as real-time nanoscale laser processing at desired locations with nanoscale positional accuracy, biomedical applications, high-resolution diagnostic tools, and semiconductor manufacturing [27-30].

In this paper, we report a unique super-resolution instrument SuperNANO with dual functions of super-resolution imaging and fabrication, which realizes simultaneous label-free super-resolution imaging and direct laser nanofabrication for the first time. We present the versatility of our instrument by demonstrating its proficiency in creating anti-counterfeiting security markings on an aluminium film sample. Through real-time direct laser writing, we generate 'invisible' nano-security features (nanoscaled) with arbitrary shapes at precise locations on the sample. We demonstrate applications in two-level anti-counterfeiting marking for coveted security features, as well as nanoscissors precise directional cutting of silver nanowires with a width of 80 nm based on the synchronized nanoimaging system. In our experiments, the gap distance between the sample and lens was monitored by an auxiliary side-view microscope system which turned to be critical for achieving reliable distance control in instrumentation. Compared with AFM-microsphere combined dual function system.

## 2. SuperNANO System Design

*2.1. SuperNANO System Setup*

Figure 1a shows the schematic of the SuperNANO system. A Thorlabs QSL103A Picosecond Microchip Laser (wavelength of 1030 nm, with a pulse duration ranging from 550 ps ± 100 ps, repetition rate of 100 kHz) is the core of the system. This laser undergoes meticulous control as it passes through a series of essential components. Initially, it encounters a precision shutter and a beam expander, both meticulously managed to ensure optimal beam quality and intensity. Subsequently, the laser enters the G3 Base galvo-scanner from BEIJING JCZ TECHNOLOGY CO., a dynamic component capable of achieving scanning speeds of up to 5000 mm/s. This swift scanning capability facilitates precise nanopatterning on the target substrate. The setup of the SuperNANO system is partitioned into distinct left and right sections, each dedicated to specific tasks. On the left, the focus is on super-resolution fabrication, while the right segment caters to super-resolution imaging.

These sections were combined in the middle, seamlessly sharing a common optical pathway and a pivotal component known as the unibody superlens objective (USO). This integration ensures efficient utilization of resources and facilitates seamless transitions between fabrication and imaging processes. A 50:50 beam splitter plays a role in dividing the laser beam, allocating portions for both power metering and USO operations. This division facilitates nano-imaging capabilities by directing a portion of the laser beam towards reflection and subsequent capture by a CCD camera. To further enhance precision, the system incorporates XYZ motorized nano-stage and micro-stage mechanisms, orchestrating the alignment between the USO and the substrate. Meanwhile, the illuminations are carefully curated, featuring an LED illumination and an aperture diaphragm to ensure optimal lighting conditions for nano-scale manipulation. Furthermore, the system boasts sophisticated monitoring and control functionalities, exemplified by the inclusion of a side-viewing microscope seamlessly integrated with the stage system. This combination enables real-time monitoring and precise adjustment of gap distances, further enhancing the system's capabilities in achieving unparalleled levels of precision and control in nanoscale manipulation.

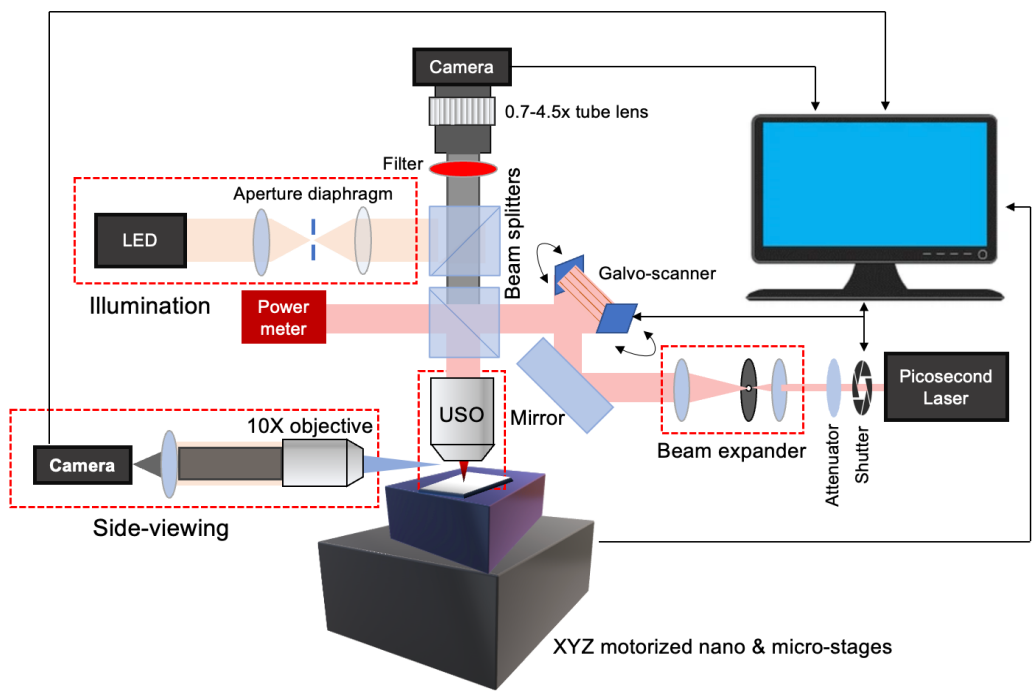

**Figure 1.** Schematic of the SuperNANO system setup.

*2.2. Fabrication of USO*

Figure 2a shows the manufacturing process of the USO lens (Fig. 2b), which is based on integrating a conventional objective (with a numerical aperture of 0.65) with an 80 μm BaTiO3 (BTG) microsphere lens in a single design, as first introduced in our previous work [31]. The two components of USO are mechanically bonded together through a transparent encapsulation layer, typically composed of ultraviolet (UV) glue or polydimethylsiloxane (PDMS). The fabrication process commences with the attachment of a microsphere onto a UV glue-coated objective (Fig.2-a1), facilitated by the precise manipulation of the Z-axis of the nano-stage to position the microsphere in touch and detach from the substrate surface (see Fig. 2 a2 and a3). Subsequently, the microsphere is secured in place through UV light curing, facilitated by the side-viewing component of the system.

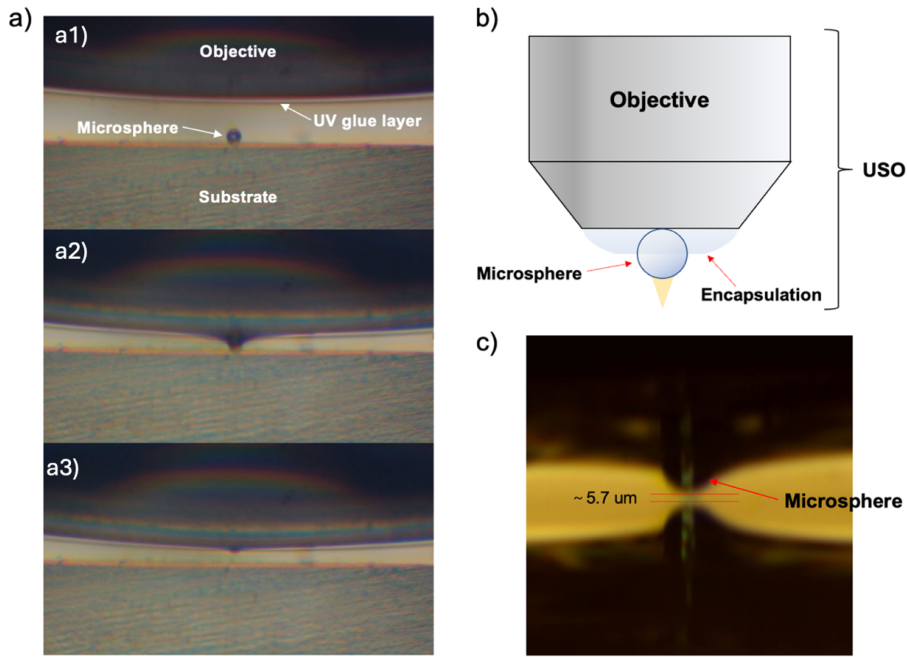

**Figure 2.** a) Production process of USO objective made by microsphere and Plano-Convex lens: a1) objective with UV glue approaching substrate, a2) touching the substrate, a3) detaching from the substrate, b) Unibody Superlens Objective, c) Focusing distance measurement by the side-view microscope system. The substrate acts like a mirror in the image.

In our work, the BTG microsphere used has a refractive index of 1.95, and the UV glue (NOA 60) has a refractive index of 1.52 at visible wavelengths. For super-resolution applications, a microsphere size of several to several tens of micrometers is recommended in the literature. Here, we have chosen a large-end size of 80 μm to ensure a working distance of about 5-6 μm for patterning applications. Smaller BTG microspheres would result in a smaller working distance, making fast and reliable nano-patterning more challenging [32-34]. Furthermore, the encapsulation status of the microsphere—whether fully or partially encapsulated within the bonding layer—provides additional control over the superlens performance. Partial encapsulation, for instance, may allow for nuanced adjustments to the microsphere's position and orientation, potentially fine-tuning the focal point and enhancing the overall resolution. In short, the final focusing resolution is mainly controlled by the microsphere size, material, and encapsulation status (e.g., partial or full encapsulation).

Figure 2c depicts the measurement of the working distance of the USO objective focusing as captured by the side-view microscope system. By adjusting the Z-axis of the nano-stage, the focal length of the fabricated USO is determined to be 5.7 ± 0.03 μm. Here, the microsphere lens has a mirrored image generated by the underlying reflective substrate. This side-view imaging system not only enables accurate focal length measurements but also effectively monitors the position of the superlens, playing a crucial role in safeguarding both the lens and the sample.

*2.3. Muti-level Laser Security Marking*

Counterfeiting is a global issue and has considerable negative impacts on our society, both economically and socially. Laser marking has been used as an effective tool in anti-counterfeiting applications. Compared to ink-based marking, laser-marked patterns are difficult to rub off, making them an ideal solution for traceability and anti-counterfeiting purposes. Laser marking creates a permanent and indelible mark in various forms including text, image, and 2D codes (QR code, Data Matrix Code and DotCode) [26,27]. However, with the rapid growth of laser technology, especially Fiber laser technology, the prices of conventional laser marking systems have considerably dropped in recent years, which has led to an increase in counterfeiting activities. Therefore, we need new innovations to enhance the security level of our markings, preventing counterfeiters with laser marking systems from copying or reproducing our fabrications. Here, our solution is multi-level security marking: the sample will be first marked with microscale markings, then a second level of nanoscale marking will be added by using the developed SuperNANO system. More levels of security can be realized by encoding nanoscale information in different strategies. Compared to other nanofabrication techniques such as photolithography, our technique does not involve complex processes such as spin coating, exposure and lift off, but a single step of direct laser writing. When coupled with ultrafast laser sources, the technique can be performed on almost any materials surface or inside body of transparent materials [36,37].

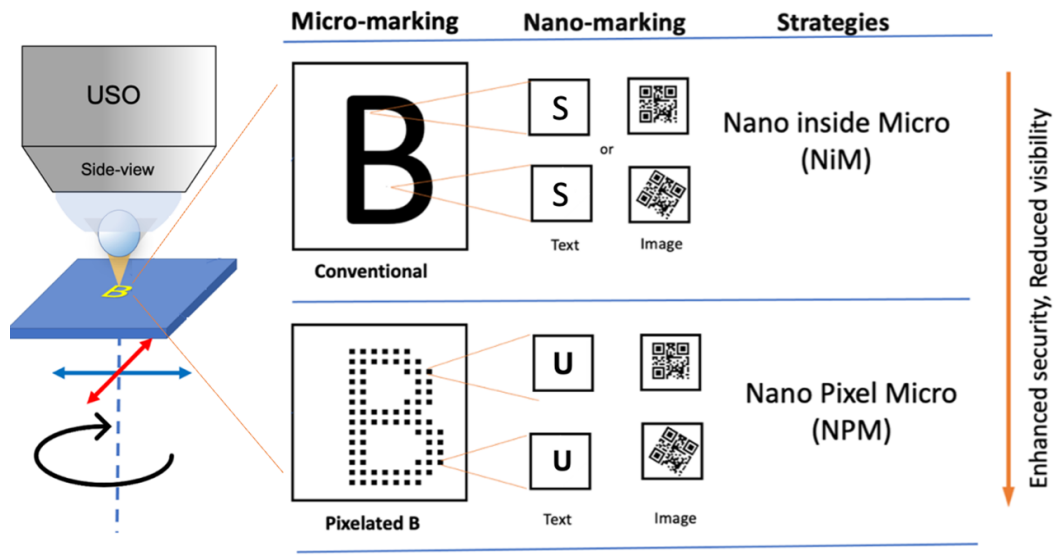

**Figure 3.** Strategies of multi-level laser security marking. NiM (Nano inside Micro): On top of conventional laser marking, adding designed nano-pattern inside. NPM (Nano Pixel Micro): Pixelize micro-marking, replace each pixel with designed nano-pattern.

Figure 3 shows two proposed multi-level security marking strategies. The first is the Nano inside Micro (NiM) strategy, where nanoscale markings (the letter 'S') were added to a microscale marking (the letter 'B'). The second strategy is Nano Pixel Micro (NPM), where the conventional laser marking design was first pixelized (see the pixelized 'B' in Figure 3). Each pixel is then replaced with a nano-design (the letter 'U'), either text or an image. Each pixel is laser nano-marked, and the overall fabricated image forms a micro marking image. Compared to the NiM strategy, the pixelized mark has reduced visibility but improved security.

## 3. Results and Discussion

### 3.1. NiM Strategy Marking

Figure 4 showcases the advanced capability of the SuperNANO system in achieving nanoscale-resolution marking within microscale structures. The image displays a large 'B' letter, microscopically engraved on the surface of an aluminum sample. Zooming into the image, a smaller 'S' letter can be observed within the structure of the 'B'. In the magnified view, the dimensions of the nanoscale marking become evident, with the 'S' letter measuring approximately 330 nm in line width. This 'S' letter is etched with nanoscale precision, demonstrating the remarkable accuracy and control of the SuperNANO system in creating intricate patterns on a nano level.

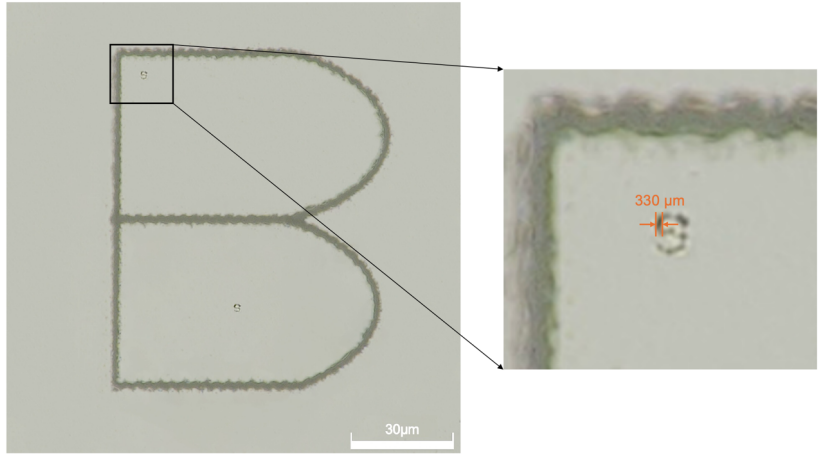

**Figure 4.** Microscope image of laser marked microscale 'B' letter with inside nanoscale 'S' letter.

*3.2. NPM Strategy Marking*

Then we demonstrate the ability of nanoscale-resolution marking by the SuperNANO system—a microscale 'U' letter marked on the aluminum sample surface with nanoscale marking resolution, as shown in Figure 5. Each 'U' exhibits dimensions of 1.32 μm in length and 1.98 μm in width, with a marking resolution of 320 nm represented by the line width. Dark field illumination in Fig. 5(b) emphasizes surface irregularities and variations in reflectivity by illuminating the sample from oblique angles, thereby enhancing the visibility of markings and making them stand out against the background. These irregularities are mainly caused by the small mechanical vibrations and laser energy fluctuations associated with the system. Moreover, the marking process is seamlessly integrated with the imaging system, providing real-time observation and monitoring of the fabrication process.

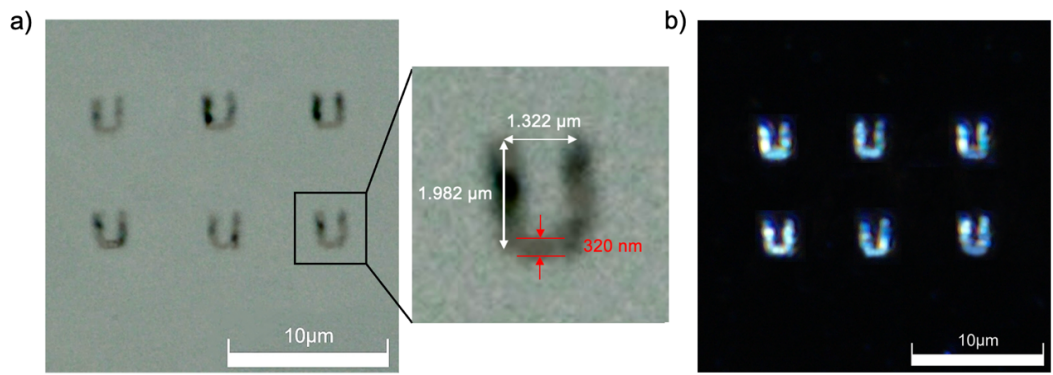

**Figure 5.** Microscope image of laser marked 'U' letters under a) bright field and b) dark field imaging modes.

Figure 6 demonstrates two examples of NPM strategy marking. In the first one, Fig. 6(a), the pixelated micro-sized letter 'B' is formed by nano-sized elements of 'U' with a line width of 330 nm, where BU represents Bangor University. The pixelated nano element offers versatility, allowing for the substitution of any desired letter form. As demonstrated in the second example in Figure 6(b), the original pixel unit 'U' is replaced with 'S', forming BS representing Bangor Superlens, while maintaining a consistent line width of approximately 340 nm. Notably, the spacing between each letter, referred to as the pixel unit, is adjustable, affording precise control over the resolution of the entire pixelated letter 'B'. The spacing was set to 2 μm and 3 μm in Fig. 6(a) and Fig. 6(b), respectively. In the system, the nano-letters 'U' or 'S' are generated by a Galvoscanning beam, while the position of each letter is controlled by a micro-stage. This setup enables not only the creation of intricate pixelated structures but also offers flexibility in adjusting resolution and letter forms, making it highly adaptable to diverse nano-marking applications.

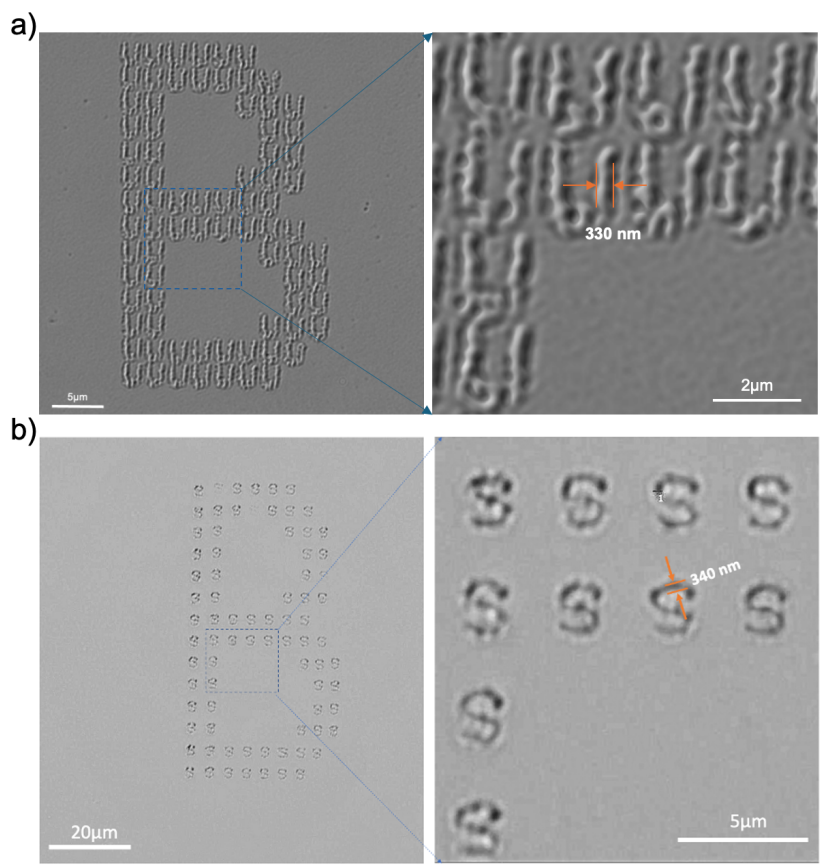

**Figure 6.** Microscope image of a) pixelated 'B' with each pixel is marked with a nano letter 'U', b) pixelated letter 'B' with each pixel is marked with a nano letter 'S'.

*3.3. Nanowire Cutting*

To explore how the resolution limit in manufacturing we performed nano-cutting experiments by the SuperNANO system on a silver nanowire (AgNW) which has many potential applications such as transparent conductive films, photovoltaics and wearable electronics [38–40]. The AgNW was purchased from ACS MATERIAL.

As illustrated in Figure 7, the SuperNANO imaging system provides a comprehensive view of both the original and cut states of silver nanowires, showing the nanowire width around 80 nm and cutting width of approximately 440 nm. The used laser fluence is about 800 mJ/cm$^2$. Notably, the outline of the USO superlens is prominently visible as the circular edge shown in Figure 7, delineating the area available for imaging and laser marking. Within this white region, measuring approximately a circle with diameter of 10 microns, lies the focal point where precise imaging and laser marking operations can be executed. This dual-functionality system seamlessly integrates imaging and marking capabilities, thereby facilitating the accurate localization of target nanowires and enabling uniform cutting at desired positions.

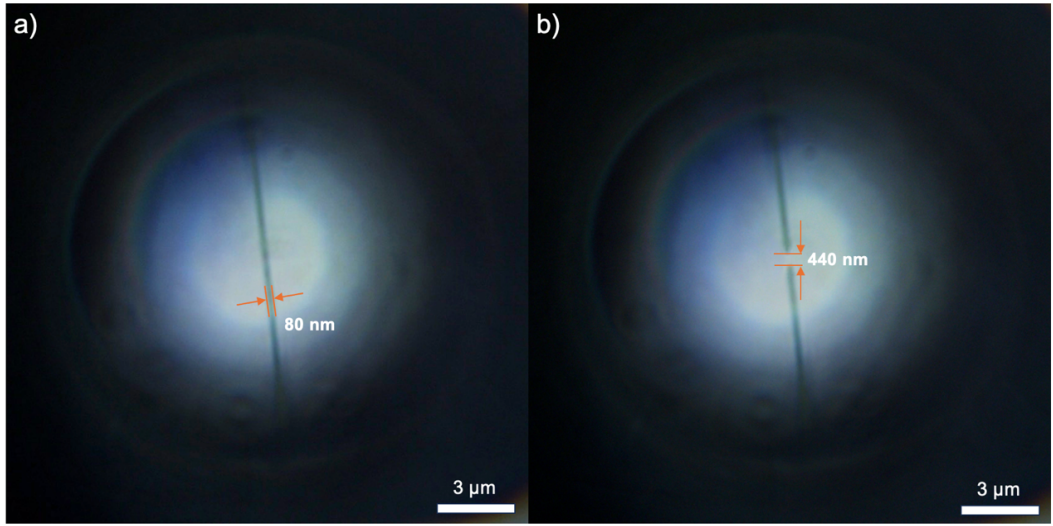

**Figure 7.** Example of silver nanowire cutting by SuperNANO system, a) before and b) after cutting.

By adjusting laser fluence, we are able to control the cutting size from 50 nm to 280 nm, when laser fluence ranging from 573 mJ/cm² to 685 mJ/cm². Figure 8 shows the cutting sizes of 50 nm and 280 nm, demonstrating its cutting ability.

Of particular significance is the achievement of the smallest cutting width approaching less than 50 nm, underscoring a significant milestone in the capabilities of our system. This breakthrough opens up avenues for precise nanoscale manipulation with unprecedented accuracy. In subsequent stages of our research, we intend to transition to a shorter wavelength femtosecond laser, aimed at enhancing the overall stability and success rate of the cutting process. Furthermore, it is noteworthy that gaps smaller than 50 nm are only distinctly recognizable under scanning electron microscope (SEM) observation, highlighting the need for continued refinement in imaging resolution. As part of our ongoing efforts to enhance system performance, we are exploring the possibility of replacing the existing components with a higher performance $TiO_2$ superlens. This strategic upgrade holds promise for further enhancing the resolution and capabilities of the SuperNANO imaging system, thereby unlocking new possibilities in nanofabrication and nanoengineering.

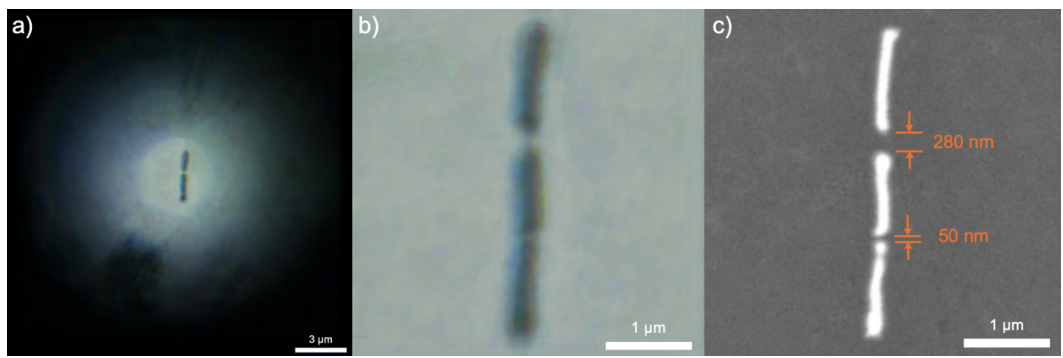

**Figure 8.** Silver nanowire cutting samples under, a) SuperNANO imaging system, b) microscope, c) SEM.

Another compelling application within our research scope is the automation of nanowire ruler cutting. This process involves the utilization of a programmable micro-stage to meticulously position and cut the nanowires at specified intervals, yielding ruler-shaped nanostructures. As showcased in the results depicted in Figure 9, obtained through both microscope and SEM measurements, this automated approach offers a systematic and controlled means of fabricating nanowire-based rulers with precision and consistency. Central to the success of nanowire ruler cutting is the management of the cutting width during the cutting process. By carefully adjusting the laser energy levels, we maintain the cutting width within the range of 850-900 nm. While opting for higher energy levels may result in larger cutting widths, it is crucial for ensuring the stability and reliability of the entire automated cutting procedure.

Moving forward, our research endeavors will be dedicated to further enhancing the stability and precision of the automated cutting process. One promising avenue involves the transition to a shorter wavelength femtosecond laser, which is anticipated to offer superior stability and control over the cutting process. By harnessing the advanced capabilities of such laser technology, we aim to achieve even smaller cutting sizes in nanowire ruler cutting. In literature, there are also reports on micro lens array (MLA) assisted laser nanopatterning. These demonstrations are, however, based on nonlinear effect in fs laser ablation and special materials being used. No super-resolution associated with MLA focusing and they are not suitable for super-resolution imaging application. Other reports on microsphere-based laser patterning technique assisted by laser trapping or AFM positioning were also considered [41], they are slow and complex so won't meet general needs for a fast and reliable nanopatterning; they are also difficult to achieve dual function as demonstrated in our instrument.

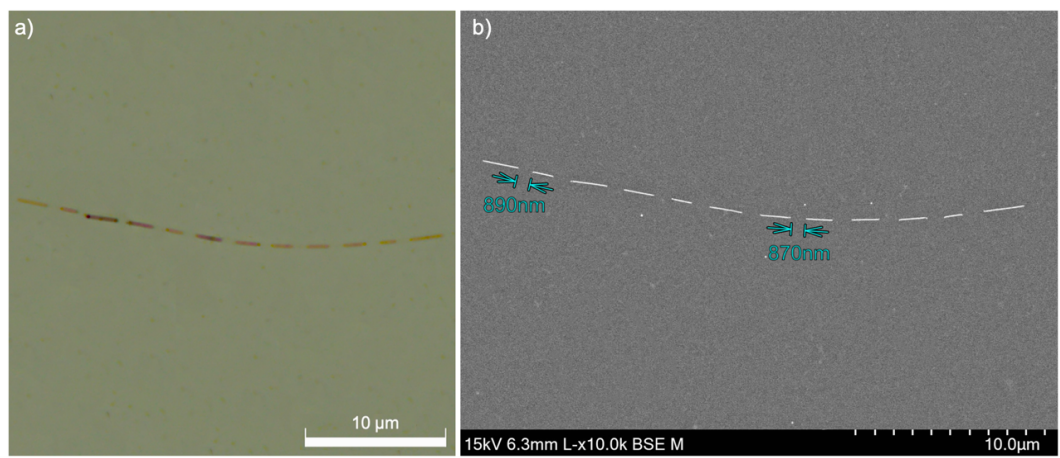

**Figure 9.** Silver nanowire ruler cutting sample under a) microscope & b) SEM.

In our quest for achieving finer features beyond the 50-300 nm threshold, we are exploring the adoption of either a higher performance superlens or a femtosecond laser with a shorter wavelength. One promising avenue involves the design and fabrication of a novel TiO2 composite superlens, leveraging $TiO_2$ nanoparticle stacking to replace the current BTG superlens. Our recent investigations have revealed that the $TiO_2$ superlens consistently outperforms the BTG microsphere superlens across various metrics including focusing spot size, imaging contrast, clarity, field of view, and imaging resolution [42]. However, the $TiO_2$ composite superlens may exhibit a shorter working lifespan compared to the BTG lens due to its composite nature.

**4. Conclusion**

In conclusion, the development of the SuperNANO instrument represents a significant advancement in nanotechnology. The demonstrated capabilities of the SuperNANO instrument extend beyond mere imaging and fabrication, encompassing applications such as two-level anti-counterfeiting marking and precise directional cutting of silver nanowires. The instrument utilizes innovative superlens technology and meticulous system design to achieve outstanding resolutions of 320 nm and 50 nm in nano security marking and nanowire cutting, respectively. The incorporation of a nano-imaging system plays a pivotal role, enabling real-time positioning, visualization, and monitoring throughout the entire marking and fabrication process. The synergistic integration of super-resolution imaging and fabrication functionalities opens up a plethora of opportunities in various fields, including nano trapping, sensing, welding, drilling, and signal enhancement and detection. Its unique combination of imaging and fabrication functions paves the way for groundbreaking advancements in a wide array of disciplines, driving innovation and pushing the boundaries of what is achievable in nanoscience and nanotechnology.

**Acknowledgments**

Leverhulme Trust (RF-2022-659), Royal society (IEC\R2\202178), Bangor University (BUIIA-S46910)